
\hoffset0.cm
\voffset0.cm
\tolerance=2000
\hbadness=2000
\vbadness=10000
\parindent=20.pt

\font\bigfett=cmbx10 scaled\magstep2

\def\pslash{\not{\hbox{\kern-2.3pt $p$}}}
\def\qslash{\not{\hbox{\kern-2.0pt $q$}}}

\magnification=\magstep1
\hsize 13cm
\baselineskip=0.22in plus .5pt minus .5pt
\parskip=15.pt
\rightline{MZ-TH/92-16}
\rightline{TTP92-17}
\rightline{ April 1992 (first version)}
\vskip 0.8cm
\centerline {\bigfett ON GAUGE INVARIANCE OF }\bigskip
\centerline {\bigfett BREIT-WIGNER PROPAGATORS}\bigskip
\bigskip\bigskip\bigskip\bigskip
\centerline {{\bf M.~Nowakowski$^{(a),}$} \footnote{$^1$}{Present address:
Physical Research Laboratory, Navrangpura, Ahmedabad 380 009, INDIA} {\it and}
{\bf A.~Pilaftsis$^{(b),}$} \footnote{$^2$}{Address after 1,
Oct.~1993,
Rutherford Appleton Laboratory, Chilton, Didcot, Oxon, ENGLAND.
E-mail address: pilaftsis@vipmza.physik.uni-mainz.de}}
\bigskip
\centerline {$^{(a)}$Inst.~f\"ur Theoretische Teilchenphysik, Universit\"at
Karlsruhe, 7500 Karlsruhe, {\it FRG}}
\bigskip
\centerline {$^{(b)}$Inst. f\"ur Physik, Johannes-Gutenberg Universit\"at,
6500 Mainz, {\it FRG}}
\bigskip\bigskip
\centerline{ Final version to appear in {\it Z.~Phys.~C}}
\bigskip  \bigskip
\centerline{\bf ABSTRACT} \smallskip
We present an approach to bosonic ($Z^0, W^{\pm}$) as well as
fermionic (top-quark) Breit-Wigner propagators which is consistent with
gauge invariance arguments. In particular, for the $Z^0$-boson propagator
we extend previous analyses and show that the
part proportional to $k_{\mu} k_{\nu}/M^2$ must be modified
near the resonance.
We derive a mass shift which agrees with results obtained
elsewhere by different methods. The modified
form of a resonant heavy fermion propagator is also given.
\vfill\eject
The quest for the analytical form of the Breit-Wigner (BW)
propagator consistent with gauge invariance has been the subject of numerous
studies  in the Standard Model (SM) [1-7]. Clearly its relevance nowadays lies
mainly in extracting the
$Z^0$ mass and width parameters from the resonant $e^+e^-$ cross section.
But the study of the $W^{\pm}$ and top propagator (we asssume $m_t > M_W+m_b$
 [8])
is not less important for instance in $e^+e^- \to W^* W^*$ [9,15].
In addition it
has been shown that the complex phase in the BW~propagator can induce new
CP-violating phenomena [10-14] which can be sizeable in SM and hence can
play an important role in search for new signals beyond SM.
This applies both to bosonic [12-14] as well as fermionic [10-11]
BW~propagators. Note, however, that the fermionic
BW~propagator is of quite different nature than its bosonic counterpart [10].
\smallskip
Recently it has been shown that the part proportional to
$k_\mu k_\nu$ tensor of the BW propagator
for the $W^{\pm}$ boson must be modified in order to satisfy $U(1)$ gauge
invariance [2]. In the present note we
prove that the same modification must take place in the resonant
$Z^0$ propagator. In fact, we get a mass shift
(i.e. the difference between $M_Z$ and $M_R$ where $M_R$ is the
renormalized mass entering the equation $(q^2-M_R^2+iq^2\Gamma_Z/M_Z)^{-1}$)
which is comparable with results obtained, for instance, in [5]. Our approach
may not be regarded only as an extension of previous analyses~[3,5,6], which
do not consider the $k_\mu k_\nu$ part of the $Z^0$ BW propagator, but
should also serve as an additional verification of them.
Finally, with the help of Ward identities
we derive the analytical form of the resonant top-quark propagator and
point out its relevance for future experiments.
\smallskip
It should be stressed here that what we have in mind is not the full
propagator, but only its resonant part. The "correctness" of our propagator
should be understood in the context of resonant processes where the usage
of the modified BW propagator leads consistently to gauge invariant results.
\smallskip
The commonly used expressions for bosonic and fermionic BW~propagators
are, respectively
$$\Delta_{\mu \nu}^V(q^2)\ = { -g_{\mu \nu} + q_{\mu}q_{\nu}/M_V^2
\over q^2-M_V^2 + iM_V\Gamma_V}\eqno(1)$$
$$S_F(\qslash)\ = \ {\qslash+ m_t \over q^2 - m_t^2 + im_t\Gamma_t}
\eqno(2)$$
where $V=Z^0, W^{\pm}$. The crucial step in deriving the aforementioned
modification of (1) for $V=W^{\pm}$ is to prove that the following
electromagnetic Ward identity [2]
$$k_{\mu}\ \overline{\Delta}^{\delta \rho}(p')\ \Gamma_{\rho \beta}^{\mu}
(p,p',k)\ \overline{\Delta}^{\beta \alpha}(p)= e \left[
\overline{\Delta}^{\delta \alpha}
(p')-\overline{\Delta}^{\delta \alpha}(p) \right]\eqno(3)$$
where
$$\Gamma_{\rho \beta}^{\mu}(p,p',k)\vert_{WW\gamma}=e \left[
(p+k)_{\rho}
g^{\mu}_{\ \beta}+(p'-k)_{\beta} g^{\mu}_{\ \rho}-(p+p')^{\mu}
g_{\rho \beta} \right] \eqno(4)$$
is the $W^+_{(\beta)}(p) \to W^+_{(\rho)}(p') + \gamma_{(\mu)}(k)$
vertex at tree level, is verified by inserting
$$\overline{\Delta}_{\mu \nu}^W(q)\ =\ {-g_{\mu \nu}+ q_{\mu}q_{\nu}
/\left(M_W^2-iM_W\Gamma_W \right) \over q^2 - M_W^2 +iM_W \Gamma_W}
\eqno(5)$$
and not the form given in (1). The authors of [2] have checked that the
process [15]
$$e^++e^- \to W^{+*} + W^{-*} + \gamma \eqno(6)$$
with the subsequent decays $W^{\pm} \to f \overline{f'}$ preserves $U(1)$
gauge invariance only if $\overline{\Delta}_{\mu \nu}^W(q)$ of (5) is used.
\smallskip
We mention that though the above considerations make use of the unitary
gauge it is evident that they can be extended to an arbitrary $\xi$ gauge.
It is remarkable that in this case we have to make the effective replacement
$$M_W^2 \ \to \ M_W^2-iM_W \Gamma_W\eqno(7)$$
not only in the $q_{\mu}q_{\nu}$ terms of the propagator, but also
in the vertices, for instance in the $\chi-W-\gamma$ vertex ($\chi$
is the charged Goldstone boson)
$$\Gamma^{\mu \nu} \vert_{\chi W \gamma}=ieM_W g_{\mu \nu} \ \to \
ie \sqrt{M_W^2 - iM_W \Gamma_W}\  g_{\mu \nu}\eqno(8)$$
where the square root is to be taken on the first sheet. It will turn out
that for the $Z^0$ and top propagator the Feynman rules must undergo a
similar effective modification when one deals with a BW propagator
in $R_{\xi}$~gauges. However, since the Goldstone boson $\chi$
will always appear as intermediate non-asymptotic state, the above
square root cannot survive in a real calculation near the $W$ resonance.
\smallskip
Yet another way to understand the modification in (5) is to consider
the process $q\overline{q'} \to Q\overline{Q'}$.
In a $\xi$ gauge there are two amplitudes at tree
level (unitarized by a BW~propagator) which we denote by
$T_{W^+}^{\xi}(q\overline{q'} \to W^+ \to Q\overline{Q'})$ and
$T_{\chi^+}^{\xi}(q\overline{q'} \to \chi^+ \to Q\overline{Q'})$.
The gauge invariance can be checked, for instance, by
$$T_{W^+}^{\xi}\ + \ T_{\chi^+}^{\xi}\ {\buildrel ! \over =}\ T_{W^+}
^{\xi \to \infty}\eqno(9)$$
This equation holds only if in $T_{W^+}^{\xi \to \infty}$ the resonant
BW~propagator is of the form given in (5). There are some subtleties
concerning the replacement (7). This and the explicit prove of our
statement we will, however, demonstrate not for the reaction
$q\overline{q'} \to Q\overline{Q'}$,
but for a similar process with an intermediate $Z^0$ instead of $W^{\pm}$.
An immediate advantage is then that we will get a modified gauge-invariant form
for the $Z^0$ propagator.
\smallskip
Consider, as before, the two amplitudes of the process $e^+e^- \to
\nu_e \overline{\nu}_e$ with massive Dirac neutrinos
$$\eqalign{&T_{Z^0}^{\xi}=T_{Z^0}^{\xi}\left(e^+e^- \to Z^0 \to \nu_e
\overline{\nu}_e
\right)=T_{Z^0}^t\ + \ T_{Z^0}^{l,\xi}\cr
&T_{\chi^0}^{\xi}=T_{\chi^0}^{\xi}\left(e^+e^- \to \chi^0 \to \nu_e
\overline{\nu}_e\right)\cr}\eqno(10)$$
In (10) we have split the $Z^0$ amplitude in components proportional to
$g^{\mu \nu}$ (denoted by $t$) and $k_\mu k_\nu$ terms (denoted by $l$).
Note that it is not essential to consider
massive Dirac neutrinos. This process is chosen as we can
demonstrate our points easily in this case. We could equivalently consider
the process, say $e^+e^- \to \mu^+ \mu^-$, and for the sake of simplicity
decouple the photon by putting $\sin \theta_W=0$.
\smallskip
Our claim is now that in order to fulfill the equation
$$T_{Z^0}^{\xi}\ + \ T_{\chi^0}^{\xi}\  {\buildrel ! \over =}\ T_{Z^0}^{\xi
\to \infty}\eqno(11)$$
it is necessary that the $Z^0$ BW~propagator takes the same form as
the $W$ propagator in (5) with the obvious replacements $M_W, \Gamma_W
\to M_Z, \Gamma_Z$. Furthermore in complete analogy to (7) we have to make
the shift $M_Z^2 \to M_Z^2-iM_Z\Gamma_Z$ in the vertices.
To show this we introduce in the
first step four different mass parameters. The first one just defines the
position of the pole
$$\tilde{M}_Z^2\ =\ M_Z^2 \ -\ iM_Z\Gamma_Z \eqno(12)$$
Very often one uses also the definition
$$\tilde{M}_Z^2\ =\ \left(M_Z \ -\ i\Gamma_Z/2 \right)^2 \eqno(13)$$
Both definitions are matters of convention.
The only physical model independent quantity in a BW~propagator is the
complex pole position [3,5,6], i.e. $q^2_{pole}=\tilde{M}^2_Z$, since this
is a fundamental property of the S--matrix theory.
For more details we refer the reader to [7].
\smallskip
The second mass parameter enters the longitudinal part of the $Z^0$-propagator
$$\overline{\Delta}_{\mu \nu}^Z(q) \ =\ {-g_{\mu \nu}\ +\ q_{\mu}q_{\nu}/(
M_Z^{'})^2
\over q^2 \ -\ \tilde{M}_Z^2}\eqno(14)$$
The remaining masses appear in the $\chi^0$ propagator and in the product
of coupling constants of the pseudo-Goldstone boson $\chi^0$  to fermions.
$$\eqalign{&\Delta_{\chi^0}(q)\ =\ {1 \over q^2 \ -\ \xi (M_Z^{''})^2}\cr
&g_{\chi^0 \nu \overline{\nu}} \cdot g_{\chi^0 e \overline{e}}\ =\
-{e^2 \over 4 \sin^2 \theta_W \cos^2 \theta_W (M_Z^{'''})^2} \cr}\eqno(15)$$
It is sufficient to concentrate on the $\xi$-dependent part which can be
cast into the form
$$\eqalign{
T_{Z^0}^{l,\xi}\ +\ T_{\chi^0}^{\xi}\ & \propto \ {1\ -\ \xi \over
q^2 \ -\ \tilde{M}_Z^2}\ {1 \over q^2 \ - \ \xi (M_Z^{'})^2}\ +\ {1 \over
(M_Z^{'''})^2}\ {1 \over q^2 \ -\ \xi(M_Z^{''})^2}\   \cr
&=\ {1 \over (M_Z^{'''})^2}\ \left({1 \over q^2\ -\ \xi (M_Z^{''})^2}
\ - \ {1 \over q^2\ -\ \xi (M_Z^{'})^2}\right)\cr
&+\ {1 \over (M_Z^{'''})^2}\ {1 \over q^2\ -\ \tilde{M}_Z^2}\ {q^2\ -\ \xi
(M_Z^{'''})^2 \ -\ \left(\tilde{M}_Z^2\ -\ (M_Z^{'''})^2 \right) \over
q^2\ - \xi (M_Z^{'})^2}\cr
}\eqno(16)$$
The first term in (16) is independent of $\xi$ only if $M_Z^{''}=M_Z^{'}$.
It then follows that
$$\tilde{M}_Z^2 \ =\ (M_Z^{'})^2\ =\ (M_Z^{''})^2\ =\ (M_Z^{'''})^2
\eqno(17)$$
which completes our proof. Equation (14) together with (17) defines the
consistent gauge invariant expression for the $Z^0$ propagator.
The $k_\mu k_\nu$ part of this propagator receives the same modification
as in the case of the $W^{\pm}$ propagator (Eq.~(5)).
\smallskip
We would like to draw the reader's attention to one subtle point in
the derivation of the modified $Z^0$ propagator (Eq.~(14) and (17)).
It concerns the shift in the Feynman rules $M_Z^2 \to M_Z^2-iM_Z \Gamma_Z$
. Instead of (15) we could equally well write the product of the $\chi^0$-
coupling constants as
$$g_{\chi^0 \nu \overline{\nu}} \cdot g_{\chi^0 e \overline{e}}\ =\
-{e^2 \over 4 \sin^2 \theta_W M_W^2}\eqno(18)$$
where $M_Z$ does not appear explicitly. It would be then impossible to
show the $\xi$ independence of the amplitude. Hence if the amplitude has
a resonant behaviour we have to use the product of the
Goldstone bosons couplings as given in (15). Clearly the relation
$\cos \theta_W =
M_W/M_Z$ remains valid for physical boson masses at the zeroth order of
perturbation theory. We think that this
point needs a further investigation especially for a situation where both
resonant particles, $W^{\pm}$ and $Z^0$, are involved. In this note we will not
pursue this topic further. $SU(2)$ Ward identities may shed some light
on this problem (see treatment of the top propagator below).
\smallskip
Some comments concerning the status of a BW propagator are, however, in
order. Feynman rules are derived from a hermitian lagrangian and are
formulated for asymptotic states. The necessity to introduce a BW propagator
in order to regularize an s-channel singularity reflects to the fact that our
states are not asymptotic and will always appear as intermediate ones
(see also another example of a t-channel singularity
at the end of the paper; this singularity appears there for the same reasons).
In fact, after one performes the Dyson summation of self-energy insertions,
the BW propagator corresponds to {\it the first term} of a Laurent series
expansion~[3,6] of the full propagator in terms of the gauge-invariant pole
$q^2_{pole}$~[3,5,6]. On the same footing this leading term of this
expansion induces similar effective modifications
in the Goldstone-boson vertices due to Ward identities.
Obviously, these vertices cannot be
derived,  like Feynman rules, {\it directly} from the Lagrangian.
\smallskip
In a context closer to perturbation theory~[3,6] we will now
confirm the modification for the $k_\mu k_\nu$ part of the propagator near
the resonance as previously obtained by Eqs.~(14) and (17).
To make this explicit, let us write the
self-energy loop of the $Z^0$ conveniently as
$$\Pi_{\mu \nu}(q)=g_{\mu \nu}\Pi_T(q^2)+q_{\mu}q_{\nu}\Pi_L(q^2)
\eqno(19)$$
For simplicity we neglect $Z^0\gamma$ mixing. The inclusion of these mixing
effects~[17] will, however, not invalidate the discussion given below.
Then, the result of the
infinite summation of self-energy graphs in an arbitrary $\xi$ gauge
leads to the following form for the (unrenormalized) $Z^0$ propagator:
$$\eqalign{\Delta_{\mu\nu}\ & = \ \Bigg[ -g_{\mu \nu}(q^2-M^2_Z+\Pi_T) \
-\ {q_\mu q_\nu \over q^2} \Big( {1 \over \xi}(q^2-
\xi M^2_Z)+\Pi_T -q^2\Pi_L \Big) \Bigg]^{-1}\ \cr
& = {1 \over q^2-M^2_Z +\Pi_T}\Bigg( -g_{\mu \nu} + {(1-\xi-\xi\Pi_L)
q_\mu q_\nu \over q^2-\xi(M^2_Z-\Pi_T +q^2\Pi_L)} \Bigg)\cr}\eqno(20)$$
The form~(20) has a physical complex pole $\tilde{M}^2_Z$ and can be
determined by solving the  equation
$$\tilde{M}^2_Z\ -\ M^2_Z\ +\ \Pi_T(\tilde{M}^2_Z)\ =\ 0 \eqno(21)$$
which is independent of the gauge parameter $\xi$.
Setting $\tilde{M}^2_Z=M^2_Z-iM_Z\Gamma_Z$, we can identify $M_Z$ and
$\Gamma_Z$ as the physical mass and the decay width of the $Z^0$ boson,
respectively.
By expanding~(20) in Laurent series~[6,7],
it is  easy to see that the first
term of this sum takes the form of Eq.~(5) with $M_W$ and $\Gamma_W$
replaced by $M_Z$ and $\Gamma_Z$, respectively. At the end of
this note, we will make some additional remarks.
\smallskip
To compare our $Z^0$ propagator with other results (for instance [5])
let us rewrite
(14) in such a way that the width appears to be $q^2$-dependent [18].
We will retain only terms up to ${\cal O}(\Gamma_Z^2)$. First we define
$$\eqalign{&\gamma_Z\ =\ {\Gamma_Z \over M_Z} \cr
&(M_Z^R)^2\ =\ M_Z^2\ -\ {\Gamma_Z^2 \over 4}\ +\ \Gamma_Z^2 \cr}\eqno(22)$$
After some straightforward algebra one obtains
$$\overline{\Delta}_{\mu \nu}^Z(q)\ =\ {\left(-g_{\mu \nu}\ +\
{\displaystyle {q_{\mu}
q_{\nu} \over (M_Z^R)^2}}(1\ +\ i\gamma_Z)\right)(1\ +\ i\gamma_Z) \over
q^2\ -\ (M_Z^R)^2 \ +\ iq^2 \Gamma_Z/M_Z}\eqno(23)$$
where we have used the pole position defined by (13). Recall that in (23)
we have neglected terms of order ${\cal O}(\Gamma_Z^3)$ wherever possible
 i.e. we have approximated $\Gamma_Z/M' \simeq \Gamma_Z/M_Z$, $M'^2+M_Z
\Gamma_Z^2 /M' \simeq M'^2+ \Gamma_Z^2$, $\Gamma_Z(M_Z-M') \simeq 0$ with
$M'^2=M_Z^2-\Gamma_Z^2/4$.
Expanding again in $\Gamma_Z/M_Z$ we get
$$M_Z^R\ =\ M_Z \left(1\ +\ {3 \over 8} \Gamma_Z^2/M_Z^2 \ +\ ... \right)
\eqno(24)$$
which is exactly the result obtained in [5], just by using the BW
propagator of the form~(5) (similar conclusions have been
reached in [2-4,6]). \smallskip
If we had used (12) instead of (13) then the formula (23) would remain
unchanged. The only difference would be the definition of $M_Z^R$,
namely $(M_Z^R)^2=M_Z^2 + \Gamma_Z^2$. This results in slightly higher
value for $M_Z^R$ ($M_Z^R=M_Z(1+\Gamma_Z^2/2M_Z^2 +...)$).
\smallskip
Finally let us discuss the case of a resonant top propagator. Similar to the
case of $W^{\pm}$ propagator we have at any loop order
$$k_{\mu}\overline{S}_t^F(\qslash)\ \Gamma^{\mu}_{\gamma}(p,q)\
\overline{S}_t^F
(\pslash)\ =\ {2 \over 3}e \left[ \overline{S}_t^F(\pslash)\ -
\ \overline{S}_t^F(\qslash) \right]
\eqno(25)$$
{}From (25) we can deduce that the gauge invariant form for the
BW propagator which respects the
Ward identity at tree level is
$$i\overline{S}_t^F(\pslash)\ =\ {i \over \pslash \ - \ m_t\ +\ i\Gamma_t/2}
\ =\ i{\pslash \ +\ m_t \ -\ i\Gamma_t/2 \over p^2 \ -\ m_t^2 \ +\
im_t \Gamma_t \ +\ \Gamma_t^2/4}\eqno(26)$$
and not the expression given in (2). To appreciate the difference we
mention that in some extensions of SM one can have $\Gamma_t/m_t \le
0.1$ [11]. The discrepancy between $m_t^R=(m_t^2 - \Gamma_t^2/4)^{1/2}$
entering Eq.~(2) and
$m_t$ could be then of the order of $GeV$. Clearly since the top physics
is not expected to be as clean as the data on the $Z^0$ pole the use of
this shift
may not be so stringent. Yet another possible application of (26) are
CP-violating processes. This possibility has been discussed in full detail in
 [10-11].
\smallskip
It is worth noting that inserting the wrong propagator (2) in the
amplitude for the process $W^+d \to W^+bg$, for example, would violate the
gauge
symmetry even while making $SU(3)_C$ gauge transformation of the gluonic
field.
\smallskip
At the end of the dicussion of the
top propagator we again address the question on the modification of Feynman
rules analogous to (7). The gauge transformations of the $W^{\pm}$ and
$Z^0$ fields read
$$\eqalign{&W_{\mu}^+\ \to \ W_{\mu}^+ \ +\ i{\partial_{\mu} \chi^+
\over M_W}\ \cr
&Z_{\mu}^0\ \to \ Z_{\mu}^0 \ +\ {\partial_{\mu} \chi^0 \over M_Z} \cr}
\eqno(27)$$
In terms of Ward identities we then get
$$\eqalign{&{k_{\mu} \over M_W}\ i\overline{S}_t^F(\pslash_t)\
i\Gamma_{W^+}^{\mu}(p_t,p_D)\ iS_D^F(\pslash_D)\cr
&+\ i\overline{S}_t^F(\pslash_t)\ i\tilde{\Gamma}_{\chi^+}(p_t,p_D)
\ iS_D^F(\pslash_D)\cr
&=\ {g_W \over 2\sqrt{2}M_W} V_{tD}\ \left[i\overline{S}_t^F(\pslash_t)(1 +
\gamma_5)\ -\ (1 -\gamma_5)iS_D^F(\pslash_D)\right]\cr}\eqno(28)$$
for the $W^+(k)+D(p_D) \to t(p_t)$ vertex and
$$\eqalign{&-i{k_{\mu} \over M_Z}\ i\overline{S}_t^F(\pslash)\ i
\Gamma_{Z^0}^{\mu}(p,q)\ i\overline{S}_t^F(\qslash)\cr
&+\ i\overline{S}_t^F(\pslash)\ \tilde{\Gamma}_{\chi^0}(p,q)\ i\overline{S}_t^F
(\qslash)\cr
&=\ -{ig_W \over 2M_W} \ \left[ (g^t_V+T^t_3{\gamma}_5)\
i\overline{S}_t^F(\qslash)\
+\ i\overline{S}_t^F(\pslash)\ (g^t_V-T^t_3{\gamma}_5) \right] \cr}\eqno(29)$$
for the $Z^0(k)+t(q) \to t(p)$ vertex. In the above $D$ denotes down-type
quark, $T_3^t$ is the weak isospin of the top quark~($T^t_3=1/2$) and
$g^t_V$ is the vector part of the $Z^0-t-t$~coupling (i.e.
$g^t_V=1/2 - 4{\sin}^2 {\theta}_W/3$). The unphysical
Goldstone boson vertices at tree level are in our case
$$\eqalign{&\tilde{\Gamma}_{\chi^+}\ =\ {g_W \over 2\sqrt 2}\ V_{tD}\
\left[{m_t - i\Gamma_t/2 \over M_W}(1 - \gamma_5)\ -\ (1 + \gamma_5){m_D
\over M_W}\right]\cr
&\tilde{\Gamma}_{\chi^0}\ =\ -g_W\ T_3^t\ {m_t - i\Gamma_t/2 \over
M_W}\ \gamma_5 \cr}\eqno(30)$$
This demonstrates once again that the Feynman rule must be modified
consistently.
In the case of a resonant top quark we have the shift $m_t \to
m_t - i\Gamma_t/2$.
\smallskip
In summary, we have set the proper framework of the approach discussed
above which defines the relevant BW propagators and the modified vertices
by Eqs.~(5), (8), (26), (30). The next step would be to show, with
the help of the Ward identities, the gauge invariance of any resonant process.
We must, however, stress again that
any resonant process can always be expressed in terms of three gauge
invariant quantities [6,7], i.e. the complex pole position
of the BW propagator $q^2_{pole}$, the residue of the pole $R_{pole}$
and a $q^2$-dependent background term $C_{back}$ which has no poles
at $q^2=q^2_{pole}$. In general, these three parameters can be
perturbatively expanded in different orders of the relevant coupling
constant. To make contact with other analyses we emphasize that {\it our
considerations can give informations only for the quantities $q^2_{pole}$
and $R_{pole}$} (i.e.~$C_{back}=0$).
However, our approach which is also based on gauge
invariance arguments not only modifies the $k_\mu k_\nu$ part of the
propagator, but can also be viewed as an additional justification of
previous works~[3,4,6].
In this context an example would be to check the $R_{\xi}$ gauge
invariance of the amplitude
$e^+ e^- \to {\mu}^+ {\mu}^-$ at any electroweak loop level without
neglecting the fermion masses, in a manner similar to that presented in~[6]
for massless fermions.
\smallskip
We conclude with a general remark on Born level singularities in the
kinematically allowed region. Beside the well-known infrared and s-channel
singularities (e.g., like the $Z^0$ case) there are also t-channel
singularities
in the SM. Let us briefly outline what is meant by these singularities. The
t-channel singular processes are closely connected to sattering processes
with unstable particles in the initial state. Consider for example the
scattering $\nu_eW^+ \to \nu_e W^+$. The total amplitude at Born level
is the sum of a s-channel and t-channel diagram.
$$\eqalign{&A(\nu_e W^+ \to \nu_e W^+)\ =\ A_s(s,t)\ +\ A_t(s,t)\cr
&A_t(s,t)\ = \ {1 \over t\ -\ m_e^2}\ f(s,t)\cr}\eqno(31)$$
where $m_e$ is the electron mass and $f(s,t)$ is a kinematical function.
The physical boundary can be easily calculated to be
$$\eqalign{&t_{max}\ =\ {M_W^4 \over s}\cr
&t_{min}\ =\ 2M_W^2\ -\ s \cr}\eqno(32)$$
It is clear that for
$$2M_W^2 \ -\ m_e^2 \le \ s\ \le {M_W^2 \over m_e^2}M_W^2 \eqno(33)$$
the value $t=m_e^2$ lies within the physical region and one picks up
a singularity in $A_t(s,t)$. There are of course many other examples of
t-channel singular amplitudes,
even in $2 \to 3$ processes like $W^+d \to W^+bg$ [19]. One of the common
feature of these singular processes is their non-localization in space-time.
This can be seen as follows. At a time $t=t_0$ the initial $W_{(i)}^+$
decays into $e^+ +\nu_e^{(f)}$ (with a disconnected $\nu_e^{(i)}$). At
$t=t_1 > t_0$ $e^+$ recombines with $\nu_e^{(i)}$ to produce $W^+_{(f)}$
(with a disconnected $\nu_e^{(f)}$ which has already been produced in the
first step).
\smallskip
The reason for the catastrophic behaviour of $A_t(s,t)$ can
be traced partly back to the fact that unstable particles cannot be prepared as
asymptotic states. This is an old well-known problem [16,20]. However, we
cannot get rid of the t-channel singularity when we replace the troublesome
$W^+_{(i)}$ by the well-defined asymptotic state $e^+\overline{\nu}_e$
and consider instead
$3 \to 2$ or $3 \to 3$ processes like
$e^+\overline{\nu}_e \nu_e \to \nu_e W^+$.
This problem is not new and has been a subject of numerous investigations in
the days of $S$-matrix theory [21]. In the case of the $3 \to 2$ or $3 \to 3$
processes {\it this $t$-channel singularity cannot be cured even by introducing
a BW propagator}.
\smallskip
The phenomenological relevance of the t-channel singular processes is
quite limited (unless one insists on using the Equivalent Vector Boson
Approximation [22]). However, since they belong to the class of singularities
which appear at Born level they are important from theoretical point of
view. The point we would like to make is that the common origin
of all tree level singularities is the truncation of the
Feynman-Dyson expansion at certain order.
The subtle point is that this perturbation
expansion can be shown to be unitary as a whole i.e. in the form
$$S\ =\ 1\ +\ i \sum_{n=1}^{\infty}T^{(n)} \eqno(34)$$
but for a finite number of terms in (34) this is, {\it in general}, not
the case~[23].
\vskip2.cm
{\bf Note added.} After our manuscript has been submitted we became
aware of two papers [24], [25] which treat similar subject. Ref. [24]
arrives in part at similar conclusion. The authors
of [25] seem, however, to have obtained a different modification of the
$q_{\mu}q_{\nu}$
terms in the propagator. We would like to make some comments on their
result. Ref. [25] uses the following decomposition of the self-energy
tensor $\Pi_{\mu \nu}$
$$\Pi_{\mu \nu}(q)=\left(-g_{\mu \nu}+\displaystyle{q_{\mu}q_{\nu} \over
q^2}\right)R_T+\displaystyle{q_{\mu}q_{\nu} \over q^2}R_L \eqno(35)$$
After summing the absorptive parts of $R_T$ and $R_L$
($ImR_T=\epsilon_T$, $ImR_L=\epsilon_L$), they obtain the following form
for the propagator in the unitary gauge:
$$D_{\mu \nu}=\displaystyle{1 \over q^2-M_W^2+i\epsilon_T}\left[-g_{\mu \nu}
+\displaystyle{q_{\mu}q_{\nu}(q^2+i\epsilon_T+i\epsilon_L) \over
q^2(M_W^2+i\epsilon_L)}\right]\eqno(36)$$
However, some difficulties in using such an expression in
practical calculations of transition matrix elements to a given
order of perturbation theory may indeed arise. The reason is that
the functions $\epsilon_T(q^2)$ and $\epsilon_L(q^2)$ are generally
$\xi$-dependent quantities
in the off-resonance region~[3,4], as well as the derivative
$\epsilon_T'(M^2_W)$~[4], and
hence the propagator~(36) induces $\xi$-(gauge)
dependence to all orders. On the other hand, $\xi$-terms coming from the
vertices can remove the gauge dependence only to a given power of coupling
constants, leaving matrix element gauge $non$-invariant at higher orders.
Furthermore, it has been stressed by Sirlin~[3] and Stuart~[6]
that the complex pole position of the propagator
$\tilde{M}^2=M^2-iM\Gamma$
is the right expansion parameter since it is intrinsically gauge invariant.
The Laurent expansion in $\tilde{M}^2$ is also
the right procedure to obtain the BW propagator from the
full one~[6]. From Eq.~(20) (and using~(35)) one then gets in the unitary
gauge
$$\Delta_{\mu \nu}=\displaystyle{{\rm Res}\Delta_{\mu \nu} \over
q^2-\tilde{M}^2}+\ \cdots
=Z\left[ \displaystyle{-g_{\mu \nu}+q_{\mu}q_{\nu}/\tilde{M}^2 \over q^2-
\tilde{M}^2}+\ {q_\mu q_\nu \over M^4}\left( R_T'\ -\  {R_T + R_L \over
M^2} \right) \right]\ +\
\cdots \eqno(37)$$
where $Z=[1+R_T'(M^2)]^{-1}$ plays the role of the wave-function
renormalization constant. The second term of the r.h.s.~of Eq.~(37)
is a non-resonant background term of ${\cal O}(\alpha)$, while the ellipses
denote contributions of ${\cal O}(\alpha (q^2-\tilde{M}^2)/q^2)$.
We consider this as an additional check that both approaches are consistent.
Also, both propagators (Eq.~(36) and Eq.~(37))
obey the optical theorem as has been explicitly demonstrated in~[25] for
the resonant point $q^2=M^2$.

\bigskip\bigskip
{\bf Acknowledgements.} We gratefully
acknowledge valuable discussions with R.Decker, Z.W\c as, K.~Schilcher,
J.~G.~K\"orner and M.~Lavelle. We also wish to thank D.~Atwood, G.~Eilam,
R.~Mendel, R.~Migneron and A.~Soni for useful comments and criticisms.
The work of A.P.~has been supported by a
grant from the Postdoctoral Graduate College of Mainz and the work of
M.N.~by Bundesministerium f\"ur Forschung und Technologie under the Grant
Nr.~06KA757.
\vfill\eject
\centerline{\bf REFERENCES}\bigskip\bigskip
[1] R.G.~Stuart, in $Z^0$ physics, Proc.~XXVth Rencontres de Moriond,
ed.~J.T.T.~Van (editions Fronti\`eres, Gif-sur-Yvette, 1990), p.~41;
M.~Consoli, A.~Sirlin, in Physics at LEP, eds J.~Ellis and R.~Peccei
(CERN-86-02, 1986) Vol.~1, p.~63ff;
B.~W.~Lynn etal., $ibid.$, p.~115ff; D.C.~Kenedy, B.W.~Lynn, C.J.-C.~Im,
R.G.~Stuart, Nucl.~Phys.~{\bf B321} (1989) 83;
A.~Borrelli, M.~Consoli, L.~Maiani and
R.~Sisto, Nucl.~Phys.~{\bf B333} (1990) 357;
J.~Ellis, `Status of Electroweak Interactions', Joint Inter. Lepton-
Photon Symposium and Europhysics Conference on High Energy Physics, Geneva,
(1991). \smallskip
[2] G.~Lopez~Castro, J.~L.~M.~Lucio, J.~Pestieau, Mod.~Phys.~Lett.~{\bf A6}
(1992) 3679 and references therein. \smallskip
[3] A.~Sirlin, Phys.~Rev.~Lett.~{\bf 67} (1991) 2127. \smallskip
[4] A.~Sirlin, Phys.~Lett.~{\bf B267} (1991) 240. \smallskip
[5] S.~Willenbrock, G.~Valencia, Phys.~Lett.~{\bf B259} (1991) 373.
\smallskip
[6] R.~G.~Stuart, Phys.~Lett.~{\bf B262} (1991) 113;
Phys.~Lett.~{\bf B272} (1991) 353 \smallskip
[7] R.~G.~Stuart, CERN-preprint, CERN-TH. 6261/91, Workshop on High-Energy
Phenomenology, Mexico City, Mexico 1991 (QCD161, W568, 1991),
and references therein.
\smallskip
[8] The CDF Collaboration, F.~Abe et al., Phys.~Rev.~{\bf D45} (1992) 3921.
\smallskip
[9] J.~Layssac, G.~Moultaka, F.~M.~Renard and G.~Gounaris, preprint
of Montpellier University in 1991, PM/90-42.
\smallskip
[10] A.~Pilaftsis, Z.~Phys.~{\bf C47} (1990) 95.\smallskip
[11] M.~Nowakowski, A.~Pilaftsis, Mod.~Phys.~Lett.~{\bf A6} (1991) 1933.
\smallskip
[12] A.~Pilaftsis, M.~Nowakowski, Phys.~Lett.~{\bf B245} (1990) 185.
\smallskip
[13] G.~Eilam, J.~L.~Hewett, A.~Soni, Phys.~Rev.~Lett.~{\bf 67} (1991) 1979;
{\bf 68} (1992) 2103 (Comment);
R.~Cruz, B.~Grzadkowski, J.F.~Gunion, Phys.~Lett.~{\bf B289} (1992) 440.
\smallskip
[14] A.~S.~Joshipura, S.~D.~Rindani, Phys.~Rev.~{\bf D46} (1992) 3008;
D.~Atwood et al., Phys.~Rev.~Lett.~{\bf 70} (1993) 1364.\smallskip
[15] A.~Aeppli, D.~Wyler, Phys.~Lett.~{\bf B262} (1991) 125.
\smallskip
[16] M.~Veltman, Physica {\bf 29} (1963) 186. \smallskip
[17] L.~Baulieu and R.~Coquereaux, Ann.~Phys.~{\bf 140} (1982) 163.\smallskip
[18] D.Y.~Bardin, A.~Leike, T.~Riemann, M.~Sachwitz, Phys.~Lett.~{\bf B206}
(1988) 539. \smallskip
[19] M.~Nowakowski, A.~Pilaftsis, Mod.~Phys.~Lett~{\bf A4} (1989) 821;
Z.~Phys.~{\bf C42} (1990) 449. \smallskip
[20] J.~Schwinger, Ann.~Phys.(N.Y.)~{\bf 9} (1960) 169.\smallskip
[21] R.~J.~Eden, P.~V.~Landshoff, P.~J.~Olive, J.~C.~Polkinghorne,
`The Analytic S-Matrix', Cambridge University Press, Cambridge 1966.
\smallskip
[22] S.~Dawson, Nucl.~Phys.~{\bf B249} (1984) 42;
R.~M.~Godbole and S.~D.~Rindani, Phys.~Lett.~{\bf B190} (1987) 192.
\smallskip
[23] S.~N.~Gupta, `Quantum Electrodynamics', Gordon and Breach 1977, p.192.
\smallskip
[24] J.~Liu, Philadelphia preprint UPR-0525T (Sept.~1992),
Phys.~Rev.~{\bf D47} (1993) R1741. \smallskip
[25] D.~Atwood et al., Technion preprint 1992, TECHNION-PH-92-39.
\smallskip
\vfill\eject
\bye